\documentclass[epj,nopacs,fleqn]{svjour}
\smartqed
\usepackage[T1]{fontenc}
\usepackage{cite,epsfig,amsmath,graphicx,amssymb,listings,slashed}
\usepackage[colorlinks,citecolor=blue,urlcolor=blue,linkcolor=blue]{hyperref}
\journalname{Eur. Phys. J. C}

\newcommand{\madanalysis}{{\sc MadAnalysis}~5}
\newcommand{\sampleanalyzer}{{\sc SampleAnalyzer}}
\newcommand{\python}{{\sc Python}}
\newcommand{\cpp}{{\sc C++}}
\newcommand{\ie}{{\it i.e.}}
\newcommand{\be}{\begin{equation}}
\newcommand{\ee}{\end{equation}}
\def\bsp#1\esp{\begin{split}#1\end{split}}

\begin{document}
\title{Designing and recasting LHC analyses with \madanalysis}
\author{
  Eric Conte\inst{1},
  B\'{e}ranger Dumont\inst{2},
  Benjamin Fuks\inst{3,4},
  Chris Wymant\inst{5}\thanks{Present address:
   Department of Infectious Disease Epidemiology,
   Imperial College London,
   St Mary's Campus,
   Norfolk Place
   London, W2 1PG, UK.}
}
\institute{
  Groupe de Recherche de Physique des Hautes \'{E}nergies (GRPHE),
  Universit\'{e} de Haute-Alsace, IUT Colmar,
  34 rue du Grillenbreit BP 50568, 68008 Colmar Cedex, France
\and
  LPSC, Universit\'e Grenoble-Alpes, CNRS/IN2P3, 53 Avenue des Martyrs,
  F-38026 Grenoble, France
\and
  Theory Division, Physics Department, CERN, CH-1211 Geneva 23,
  Switzerland
\and
  Institut Pluridisciplinaire Hubert Curien/D\'epartement Recherches
  Subatomiques, Universit\'e de Strasbourg/CNRS-IN2P3,
  23 Rue du Loess, F-67037 Strasbourg, France
\and
  Laboratoire d'Annecy-le-Vieux de Physique Th\'{e}orique,
  9 Chemin de Bellevue, F-74941 Annecy-le-Vieux, France
}

\date{Received: date / Accepted: date}

\abstract{
We present an extension of the expert mode of the
\madanalysis\ program dedicated to the design
or reinterpretation of high-energy physics collider analyses. We
detail the predefined classes, functions and
methods available to the user and
emphasize the most recent developments. The latter include the possible
definition of multiple sub-analyses and a novel user-friendly
treatment for the selection criteria.
We illustrate this approach by two concrete examples:
a CMS search for supersymmetric partners of the top quark
and a phenomenological analysis targeting hadronically decaying monotop systems.
}

\titlerunning{Designing and recasting LHC analyses with \madanalysis}
\authorrunning{E.~Conte \textit{et al.}}

\maketitle

\vspace*{-11.5cm} \noindent 
\small{CERN-PH-TH/2014-088, LPSC 14-079, LAPTH-033/14,\\ MCNET-14-11}\\
\today
\vspace*{9.0cm}

\section{Introduction} \label{Intro}
For every experimental analysis at the CERN Large Hadron Collider (LHC),
selection criteria, widely referred to as {\it cuts}, are necessary for
the
reduction of the data-recording rate to a technically feasible level
and the discrimination between interesting and irrelevant events
for a specific physics question.
At the experimental level, it is important
to distinguish between two classes of cuts: those imposed at the trigger level,
and those imposed offline. Events failing the former are not recorded at all
and the information is lost, whereas events failing the latter are merely not
considered for the final analysis. This distinction is less
important
for the reinterpretation of an analysis based on any
sample of events other than real observed data, notably events generated by
Monte Carlo simulations of collisions to be observed assuming a
given (new) physics model. In this case, both types of cuts simply amount to conditions
on whether a given generated event is considered in the analysis or not.
However, the reinterpretation of an analysis in general requires {\it ex
novo} implementation of the full set of cuts.

Several frameworks~\cite{Conte:2012fm,Drees:2013wra,%
Barducci:2014ila} have recently been released with this
aim, based on simulated collisions
including an approximate modeling of
the detector response. Although the description of the detector
is highly simplified when compared to the full ATLAS or CMS software,
public fast detector simulations, most notably the {\sc Delphes}
program~\cite{deFavereau:2013fsa}, have been found
to provide reasonably accurate results.
Detector effects could instead be mimicked
by scale factors derived from unfolded
data as done in \textsc{Rivet}~\cite{Buckley:2010ar}. While
this is an excellent approach for Standard Model measurements,
detector unfolding is
however not yet practicable for beyond the Standard Model searches
in general.
Another alternative way, which does not rely on event simulation,
uses results published by the experimental
collaborations in the context of so-called Simplified Models
Spectra, as in the works of
Refs.~\cite{Kraml:2013mwa,Papucci:2014rja}. While much faster,
this is, however, less general.

In the present work, we focus on the expert mode of the
\madanalysis\ program~\cite{Conte:2012fm,Conte:2013mea}
dedicated to the implementation of any
analysis based on a cut-and-count flow (in contrast to analyses relying on
multivariate techniques) and the investigation of the associated effects on
any Monte Carlo event sample.
The implementation of an analysis is facilitated by the
large number of predefined functions and methods included in the
\sampleanalyzer\ library shipped with the package, but
is however often complicated in cases where one has several
sub-analyses which we refer to as \textit{regions} (with a nod to the
terms {\it signal} and \textit{control regions} commonly used in searches
for phy\-sics beyond the Standard Model). The complication arose from
the internal format handled by \sampleanalyzer, which assumed the
existence of a single region. While this assumption is convenient for
prospective studies, \textit{i.e.}, the design of new analyses,
it is rarely fulfilled by
existing analyses that one may want to recast. In order to allow
the user to both design and recast analyses, we have consequently extended the
\sampleanalyzer\ internal format to support analyses with multiple regions
defined by different sets of cuts. We have also expanded the
code with extra methods and routines to facilitate
the implementation of more complex analyses by the user.

\begin{figure}
\begin{center}
 \includegraphics[width=0.8\columnwidth]{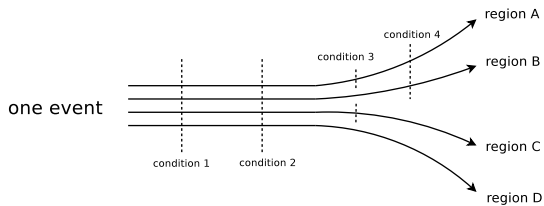}
\end{center}
\caption{Schematic illustration of the definition of different regions in
  which a given event can be counted (or not), based on different
  combinations of selection cuts.}
\label{fig:DivergingRegions}
\end{figure}

In the context of analyses which effectively contain sub-analyses, a further
useful classification of cuts can be made: namely into those which are
common/shared by different regions, and those which are not,
the latter serving to define the different sub-analyses themselves.
Figure~\ref{fig:DivergingRegions} schematically illustrates an
analysis containing four regions, which are defined by two
region-specific cuts imposed after two common cuts.
Some thought is required concerning
the best way to capture in an algorithm the set
of selection requirements shown in Figure~\ref{fig:DivergingRegions}.
For the common cuts (cuts 1 and 2 on the figure) this is clear: if the
selection condition is failed, the event is vetoed (\ie, we ignore it and
move on to analyzing the next event). Thereafter we have two conditions
to check (cuts 3 and 4), but they apply to
different regions. In terms of pseudo-code the most obvious, although
not the most efficient, method for implementing these third and
fourth cuts is
\begin{verbatim}
 count the event in region D
 if (condition 3)
 {
   count the event in region C
   if (condition 4)
   {
     count the event in region A
   }
 }
 if (condition 4)
 {
   count the event in region B
 }
\end{verbatim}
One important drawback of this naive approach is the duplication of
the check of the fourth condition. In the simple style of implementation
of the cuts above, this is unavoidable: condition 4 must be checked both
inside and outside the scope of condition 3. With the two region-specific
cuts that we have here, there is only one such clumsy duplication present in
the code. However as the number of such cuts grows, the situation rapidly
gets worse. For instance, considering growing the decision tree shown in
Figure~\ref{fig:DivergingRegions} to include $N$ region-specific cuts,
combined in all possible permutations to define $2^N$ regions would
deepen the nesting of the above pseudo-code and lead to $2^N -(N+1)$
unnecessary duplications of checks. Moreover, each of those needs to be
carefully implemented by the user in the correct scope, a task becoming
less and less straightforward for large values of $N$.

Ideally the algorithm should be structured so that there is {\it no}
unnecessary duplication, which is
one of the new features of the latest version
of \sampleanalyzer, the \cpp\ core of the \madanalysis\ program. Both
can be obtained from the \madanalysis\ website,\\
  \verb+  https://launchpad.net/madanalysis5+\\
and all the features described in this paper are available
from version 1.1.10 of the code onwards. This document supersedes
the previous version of the manual for the expert mode of the program~\cite{Conte:2012fm}.

The remainder of this paper is organized as follows. In Section~\ref{sec:ma5},
we recall the basic functionalities
of \madanalysis\ for the implementation of physics analyses
in the expert mode of the program, which has been
extended according to the needs of our users. Moreover, we
introduce the new features of the \sampleanalyzer\ kernel.
Two concrete examples are then provided in
Section~\ref{sec:examples}: the first is
the reimplementation of a CMS search
for supersymmetric partners of the top quark in events with a single
lepton and missing energy~\cite{Chatrchyan:2013xna}, and the second
the design of a monotop analysis where the monotop system decays in the hadronic
mode~\cite{Agram:2013wda}.
Our work is summarized in Section~\ref{sec:conclusions}.

\section{The expert mode of \madanalysis}\label{sec:ma5}
In this section we present the manner in which physics
analyses are implemented in the expert mode of \madanalysis. We begin with
the creation of an analysis template (Section~\ref{sec:template}),
followed by
the addition of several analyses to the same template
(Section~\ref{sec:multianalyses}). We then
discuss the methods and classes allowing
effective implementation of an analysis (Section~\ref{sec:sa}),
its compilation and execution (Section~\ref{sec:exec})
and finally the structure of the output files
(Section~\ref{sec:saf}).

\subsection{Creation of an analysis template}
\label{sec:template}
In the \textit{expert mode} of the program,
the user is asked to write his/her analysis in \cpp, using all the classes
and methods of the \sampleanalyzer\ library. To begin implementing
a new analysis, the user is recommended to use the \python\ interpreter
of \madanalysis\ to create a working directory. This is achieved
by starting \madanalysis\ with the command
\begin{verbatim}
 ./bin/ma5 <mode> -E
\end{verbatim}
where the value of \texttt{<mode>} refers to
an analysis of events generated at
the parton level (\texttt{-P}), hadron level (\texttt{-H})
or reconstructed level (\texttt{-R}). It is then enough to
follow the instructions displayed on the screen -- the user is
asked for the names of the working directory and of his/her analysis, which
we denote by \texttt{name} in the rest of Section~\ref{sec:ma5}.
The directory that has been created contains
three subdirectories: the \texttt{Input}, \texttt{Output}
and \texttt{Build} directories.

Use of the \texttt{Input} directory
is optional. It has been included in the analysis template in order to
have a unique structure for both the normal and expert modes
of \madanalysis. In the normal mode, its purpose is to collect text files
with the lists of paths to the event samples to analyze.
The \texttt{Output} directory has been conceived to store the results of each
execution of the analysis. The \texttt{Build} directory includes a series
of analysis-independent files organized into several sub-directories,
together with files to be modified by the user.

At the root of the \texttt{Build} directory, one finds one {\sc bash}
script together with its {\sc tcsh} counterpart. These
scripts set appropriately the environment variables necessary
for the compilation and execution of
an analysis within the \madanalysis\ framework.
They are initiated by typing in a (\texttt{bash} or \texttt{tcsh})
shell the respective commands
\begin{verbatim}
 source setup.sh     source setup.csh
\end{verbatim}

A \texttt{Makefile} is also available so that the standard
commands
\begin{verbatim}
 make     make clean     make mrproper
\end{verbatim}
can be used to (re)compile the analysis (see
Section~\ref{sec:exec}).
The final executable is obtained from two
pieces -- a library and the main program. The library originates from
the merging of
the \sampleanalyzer\ library and the analysis of the user, and is stored
in the subdirectory \texttt{Build/Lib}. The main program is located in the
\texttt{Build/Main} subdirectory and has a simple structure.
It first initializes the analysis, then runs the analysis over all events
(possibly collected into several files) and eventually
produces the results in the \texttt{Output}
directory previously mentioned.

The \texttt{Build} directory contains moreover the \texttt{SampleA\-na\-ly\-zer}
subdirectory that stores the source and header files associated
with the analysis being implemented (\texttt{A\-na\-ly\-zer/na\-me.cpp} and
\texttt{A\-na\-ly\-zer/na\-me.h}), together with a \python\ script,
\texttt{new\-A\-na\-ly\-zer.py}, dedicated to the implementation of
several analyses into a single working directory.
The \texttt{Analyzer} subdirectory additionally includes a list with all
analyses implemented in the current working directory (\texttt{analysisList.h}).
More information about those files is provided
in the next subsections.

\subsection{Merging several analyses in a single working directory}
\label{sec:multianalyses}
In Section~\ref{sec:template}, we have explained how to create a working
directory containing a single (empty) analysis that is called, in our example,
\texttt{name}. The analysis itself is
implemented by the user in a pair of
files \texttt{name.cpp} and \texttt{name.h},
which should be consistently referred to in the file \texttt{analysisList.h}.
In addition, the main program (the file \texttt{Build/Main/main.cpp})
takes care of initializing and executing the analysis.
The structure of this analysis provides guidelines for the implementation of
any other analysis -- \texttt{newname} for
the sake of the example -- in the same working directory. This new analysis
has to be written in
the two files \texttt{newname.cpp} and \texttt{newname.h} (stored in the
\texttt{Build/SampleAnalyzer/Analyzer} directory)  and referred to in the
\texttt{analysisList.h} file.
The main program also needs to be
modified in order to initialize and execute the new analysis, in addition to
the first analysis (\texttt{name}).

All these tasks have
been automated (with the exception of the implementation of the analysis
itself) so that the user is only required to run the python script
\texttt{newAnalyzer.py} by typing in a shell the command
\begin{verbatim}
 ./newAnalysis.py newname
\end{verbatim}
from the \texttt{Build/SampleAnalyzer} directory.

\subsection{Implementing an analysis in the \madanalysis\ framework}
\label{sec:sa}
\subsubsection{General features}
As briefly sketched in the previous subsections, the implementation of a specific
analysis within the \madanalysis\ framework consists of providing the
analysis \cpp\ source and header files \texttt{name.h} and \texttt{name.cpp}.

The header file contains the declaration of a class dedicated to the
analysis under consideration. This class is defined as a child class inheriting from
the base class \texttt{AnalysisBase}, and includes, in addition to constructor
and destructor methods, three functions to be implemented by the user (in the
source file \texttt{name.cpp}) that define the analysis itself. The first
of these, dubbed \texttt{Initialize}, is executed just once prior to
the reading of the user's set of events. In particular, it enables one both to declare
selection regions and to associate them with a series of cuts and histograms. It returns
a boolean quantity indicating whether the initialization procedure has been achieved
properly. If not, the execution of the main program is stopped.
The second method, named \texttt{Execute}, is the core of
the analysis and is applied to each simulated event provided by the user.
Among others things, it takes care of the application of the selection cuts
and the filling of the various histograms. This function returns a boolean
quantity that can be used according to the needs of the user, although
it is by default not employed.
Finally, the last function, a function of void type called
\texttt{Finalize}, is called once all events have been read
and analyzed.
Moreover, the user is allowed to define his/her own set of functions
and variables according to his/her purposes.

\renewcommand{\arraystretch}{1.5}%
\begin{table*}
  \centering
  \begin{tabular*}{\textwidth}{@{\extracolsep{\fill}}p{.21\textwidth}p{.74\textwidth}@{}} \hline\hline
    \texttt{AddCut(...)} & Declares a cut and links it to a set of regions.
     A single string must be passed as an argument, corresponding to the user-defined
     name of one of the selection cuts of the analysis. If no other argument is
     provided, the cut is associated with
     all declared signal regions. Otherwise, an additional
     single string or an array of strings, corresponding to
     the name(s) of the region(s) associated with the cut, can optionally be specified.\\
    \texttt{AddHisto(...)} & Declares a histogram. The first
     argument is the name of the histogram, the second one is
     the number of bins (an integer number), the third and fourth arguments
     define the lower
     and upper bounds of the $x$-axis (given as floating-point numbers), respectively.
     The last argument is optional and links all or
     some of the declared regions to the histogram (see the \texttt{AddCut} method
     for more information on this feature).\\
    \texttt{AddRegionSelection(...)} & Declares a new region. This method takes a string,
      corresponding to a user-defined name for the region, as its argument.\\
    \texttt{ApplyCut(...)} & Applies a given cut. This method
      takes two mandatory arguments. The first is a boolean variable and
      indicates whether
      the selection requirement associated with a given cut is satisfied.
      The second argument is the name of
      the considered cut, provided as a string. The method returns \texttt{true}
      if at least one region defined anywhere in the analysis is
      still passing all cuts so far, or \texttt{false} otherwise.\\
    \texttt{FillHisto(...)} & Fills a histogram. The first argument is a string specifying the
      name of the considered histogram, and the second is a floating-point number providing
     the value of the observable being histogrammed.\\
    \texttt{InitializeForNewEvent(...)} & To be called prior to the analysis of each event
     at the beginning of the \texttt{Execute} function. This
      method tags all regions as surviving the cuts, and initializes the weight associated with
      the current event to
      the value defined by the user passed as an argument (given as a floating-point number).\\
    \texttt{IsSurviving(...)} & Takes as an argument the name of a region (a string).
      The method returns {\tt true} if the region survives all cut applied so far, {\tt false} otherwise.\\
    \texttt{SetCurrentEventWeight(...)} & Modifies the weight of the current event to a user-defined value
      passed as an argument (given as a floating-point number).\\
\hline\hline
  \end{tabular*}
  \caption{\small \label{tab:RSMfcts}Methods of the \texttt{RegionSelectionManager} class.}
\end{table*}
\renewcommand{\arraystretch}{1}%

The splitting of the analysis into regions, the application of the selection
criteria, and the filling of histograms are all controlled through the automatically
initialized object \texttt{Manager()} -- a pointer to an instance
of the class \texttt{RegionSelectionManager}. The member methods
of this class are listed in Table~\ref{tab:RSMfcts} and will be detailed
in the next subsections, in which we also
provide guidelines for the implementation of the functions
\texttt{Initialize}, \texttt{Execute} and  \texttt{Finalize}
in the \cpp\ source file \texttt{name.cpp}.

\subsubsection{Initialization of an analysis}
\label{sec:init}
When the analysis is executed from a shell, the program first calls the
\texttt{Initialize} method
before starting to analyze one or several event samples.

Prior to the declaration of regions, histograms and cuts, we first
encourage the user to include an electronic signature to
the analysis being implemented and to ask the program
to display it to the screen. Although this is neither
mandatory nor standardized, it improves the traceability of a
given analysis and provides information to the community about
who has implemented the analysis and which reference works
have been used. In particular for analyses that are being made public,
we strongly recommend including at least
the names and e-mail addresses of the authors,
a succinct description of the analysis and related experimental notes or
publications. Taking the example of the CMS stop search
in monoleptonic events~\cite{Chatrchyan:2013xna} presented in
Section~\ref{sec:examples}, an electronic signature could be
\begin{verbatim}
 INFO << "Analysis: CMS-SUS-13-011, arXiv:1308.1586"
      << " (stop search, single lepton)" << endmsg;
 INFO << "Recasted by: Conte, Dumont, Fuks, Wymant"
      << endmsg;
 INFO << "E-mails: " << "conte@iphc.cnrs.fr, "
      <<                "dumont@lpsc.in2p3.fr, "
      <<                "fuks@cern.ch, "
      <<                "wymant@lapth.cnrs.fr"
      << endmsg;
 INFO << "Based on MadAnalysis 5 v1.1.10" << endmsg;
 INFO << "DOI: xx.yyyy/zzz" << endmsg;
 INFO << "Please cite arXiv:YYMM.NNNN [hep-ph]"
      << endmsg;
\end{verbatim}
where the last three lines refer to the Digital Object Identifier~\cite{doi}
of the analysis code (if available) and the physics publication
for which this analysis reimplementation has been
developed.
The sample of code above
also introduces the \texttt{INFO} message service
of the \sampleanalyzer\ framework, which is presented in
Section~\ref{sec:msg_services}.

As already mentioned, each analysis region must be properly declared
within the \texttt{Initialize} function. This is achieved by making use
of the \texttt{AddRegionSelection} method of the
\texttt{Re\-gi\-on\-Se\-lec\-tion\-Ma\-na\-ger} class (see Table~\ref{tab:RSMfcts}).
This declaration requires
provision of a name (as a string) which serves as a unique identifier
for this region within both the code itself
(to link the region to cuts and histograms) and the output files
that will be generated by the program. For instance, the
declaration of two regions,
dedicated to the analysis of events with a missing transverse energy
$\slashed{E}_T > 200$~GeV and 300~GeV could be implemented as
\begin{verbatim}
 Manager()->AddRegionSelection("MET>200");
 Manager()->AddRegionSelection("MET>300");
\end{verbatim}
As shown in these lines of code,
the declaration of the two regions is handled by
the \texttt{Manager()} object, an instance
of the \texttt{RegionSelectionManager} class that is
automatically included with any given analysis. As a result,
two new regions are created and the program internally assigns the intuitive
identifiers
\mbox{\texttt{"MET>200"}} and \texttt{"MET>300"} to the respective regions.

Once all regions have been declared, the user can continue with the
declaration of cuts and histograms.
As for regions, each declaration requires
a string name which acts as an
identifier in the code and the output. Histogram declaration also asks for
the number of bins (an integer number) and the lower and upper
bounds defining the range of the $x$-axis
(two floating-point numbers) to be specified.
Both histograms and cuts must also be associated with one or more regions.
In the case of cuts, this finds its source at the conceptual level:
each individual region is
{\it defined} by its unique set of cuts.
In the case of histograms, this enables one to establish the
distribution of a particular observable {\it after} some region-specific
cuts have been applied.
The association of both types of objects to their regions follows a similar syntax,
using an optional argument in their declaration. This argument is either a string
or an array of strings, each being the name of one of the previously declared regions.
If this argument is absent, the cut/histogram is automatically associated with all regions.
This feature can be used, for example, for preselection cuts that are requirements
common to all regions.

\renewcommand{\arraystretch}{1.5}%
\begin{table*}
  \centering
  \begin{tabular*}{\textwidth}{@{\extracolsep{\fill}}p{.26\textwidth}p{.71\textwidth}@{}} \hline\hline
    \texttt{mc()->beamE().first} & Returns, as a floating-point number, the energy of the first
     of the colliding beams.\\
    \texttt{mc()->beamE().second} & Same as  \texttt{mc()->beamE().first} but for the second
     of the colliding beams.\\
    \texttt{mc()->beamPDFauthor().first} & Returns, as an integer number, the identifier of the
     group of parton densities that have been used for the first of the colliding beams. The numbering
     scheme is based on the \textsc{PdfLib}~\cite{PlothowBesch:1992qj} and \textsc{LhaPdf}~\cite{Giele:2002hx}
     packages.\\
    \texttt{mc()->beamPDFauthor().second} & Same as \texttt{mc()->beamPDFauthor().first} but for the second
     of the colliding beams.\\
    \texttt{mc()->beamPDFID().first} & Returns, as an integer number, the code associated with
     the parton density set (within a specific group of parton densities) that has been used for the first
     of the colliding beams. The numbering
     scheme is based on the \textsc{PdfLib}~\cite{PlothowBesch:1992qj} and \textsc{LhaPdf}~\cite{Giele:2002hx}
     packages.\\
    \texttt{mc()->beamPDFID().second} & Same as \texttt{mc()->beamPDFID().first} but for the second
     of the colliding beams.\\
    \texttt{mc()->beamPDGID().first} & Returns, as an integer number, the Particle Data Group
     identifier defining the nature of the first of the colliding beams. The numbering
     scheme is based on the Particle Data Group review~\cite{Beringer:1900zz}.\\
    \texttt{mc()->beamPDGID().second} & Same as \texttt{mc()->beamPDGID().first} but for the second
     of the colliding beams.\\
    \texttt{mc()->processes()} & Returns a vector of instances of the \texttt{ProcessFormat} class
     associated with the set of subprocesses described by the sample. A \texttt{ProcessFormat} object
     contains information about the process identifier fixed by the generator (an integer number accessed
     via the \texttt{processId()} method), the associated cross section in pb (a floating-point number
     accessed via the \texttt{xsection()} method) and the related uncertainty (a floating-point
     number accessed via the \texttt{xsection\textunderscore error()} method), and the maximum
     weight carried by any event of the sample (a floating-point number accessed
     via the \texttt{maxweight()} method).\\
    \texttt{mc()->xsection()} & Returns, as a floating-point number, the cross section in pb 
     linked to the event sample.\\
    \texttt{mc()->xsection\textunderscore error()} & Returns, as a floating-point number,
      the (numerical) uncertainty on the cross section associated with the event sample.\\
\hline\hline
  \end{tabular*}
  \caption{\small \label{tab:sampleformat}Methods of the \texttt{SampleFormat} class.}
\end{table*}
\renewcommand{\arraystretch}{1}%

As an illustrative example, the code
\begin{verbatim}
 Manager()->AddCut("1lepton");
 std::string SRlist[] = {"MET>200","MET>300"};
 Manager()->AddCut("MET>200 GeV",SRlist);
\end{verbatim}
would create two preselection cuts,
\texttt{"1lepton"} and \texttt{"MET>200 GeV"}, and
assign them to the two previously declared regions
\mbox{\texttt{"MET>200"}} and \texttt{"MET>300"}.
Although both cuts are associated with both regions, for
illustrative purposes we have shown two methods of doing this --
using the syntax for automatically linking to all regions (here
there are two) and explicitly stating both regions.
As a second example, we consider
the declaration of a histogram of 20 bins
representing the transverse momentum distribution of the leading
lepton, $p_T(\ell_1)$, in the range $[50,500]$~GeV. In the case
where the user chooses to associate it with
the second region only, the line
\begin{verbatim}
  Manager()->AddHisto("ptl1",20,50,500,"MET>300");
\end{verbatim}
should be added to the analysis code.

Finally, the \texttt{Initialize} method
can also be used for the initialization of one or several user-defined
variables that have been previously declared in the header file \texttt{name.h}.

\subsubsection{Using general information on Monte Carlo samples}

Simulated events can be classified into two categories: Monte Carlo events
either at the parton or at the hadron level, and reconstructed
events after object reconstruction.\footnote{Strictly speaking, there exists
a third class of events once detector simulation has been included. In this case,
the event final state consists of tracks and calorimeter deposits.
\madanalysis\ has not been designed to analyze those events and
physics objects such as (candidate) jets and electrons must be reconstructed prior to be
able to use the program.} Contrary to reconstructed
event samples, Monte Carlo samples in general contain
global information on the generation process,
such as cross section, the nature of the parton density set
that has been used, \textit{etc}. In the \madanalysis\
framework, these pieces of information are collected under the form of instances
of the \texttt{SampleFormat} class and can be retrieved by means
of the methods provided in Table~\ref{tab:sampleformat}.

The function \texttt{Execute}
takes, as a first argument, a \texttt{Sam\-ple\-For\-mat} object associated
with the current analyzed sample. In this way, if
the sample is encoded in the \textsc{Lhe}~\cite{Boos:2001cv,Alwall:2006yp},
\textsc{StdHep}~\cite{stdhep} or \textsc{HepMc}~\cite{Dobbs:2001ck} format,
the user may access most of the available information passed by the
event generator. In contrast, the other event formats supported by
\madanalysis, namely the
\textsc{Lhco}~\cite{lhco} and (\textsc{Root}-based~\cite{Brun:1997pa})
\textsc{Delphes 3}~\cite{deFavereau:2013fsa} format\footnote{In order
to activate the support of \madanalysis\ for the output
format of {\sc Delphes}~3, the user is requested to start
the \madanalysis\ interpreter
(in the normal execution mode of the program) and to type
\texttt{install delphes}.}, do not include
any information of this kind so that the first argument of the
\texttt{Execute} function is a null pointer. In the case where
the user may need such information, it will have to be
included by hand.

\renewcommand{\arraystretch}{1.5}%
\begin{table*}
  \centering
  \begin{tabular*}{\textwidth}{@{\extracolsep{\fill}}p{.075\textwidth}p{.88\textwidth}@{}} \hline\hline
    \texttt{alphaQCD()}  & Returns, as a floating-point number, the employed value for the
                           strong coupling constant.\\
    \texttt{alphaQED()}  & Returns, as a floating-point number, the employed value for
                           the electromagnetic coupling constant.\\
    \texttt{particles()} & Returns, as a vector of
                           \texttt{MCParticleFormat} objects, all the
                           final-, intermediate- and initial-state particles of the event.\\
    \texttt{processId()} & Returns, as an integer number, the identifier of the physical process related to
                           the considered event.\\
    \texttt{scale()}     & Returns, as a floating-point number, the employed value for the factorization scale.\\
    \texttt{weight()}    & Returns, as a floating-point number, the weight of the event.\\ \hline
    \texttt{MET()}       & Returns, as an \texttt{MCParticleFormat} object, the missing
                           transverse momentum $\slashed{\vec{p}}_T$ of the event. The particles relevant for the
                           calculation must be properly
                           tagged as invisible (see Section~\ref{sec:phys_services}).\\
    \texttt{MHT()}       & Returns, as an \texttt{MCParticleFormat} object, the missing
                           transverse hadronic momentum $\slashed{\vec{H}}_T$ of the event.
                           The particles relevant for the
                           calculation must be properly
                           tagged as invisible and hadronic (see Section~\ref{sec:phys_services}).\\
    \texttt{TET()}       & Returns, as a floating-point number, the total visible transverse energy of
                           the event $E_T$. The particles relevant for the
                           calculation must not be
                           tagged as invisible (see Section~\ref{sec:phys_services}).\\
    \texttt{THT()}       & Returns, as a floating-point number, the total visible transverse hadronic energy of
                           the event $H_T$. The particles relevant for the
                           calculation must be properly
                           tagged as hadronic, and not tagged as invisible
                           (see Section~\ref{sec:phys_services}).\\
  \hline\hline
  \end{tabular*}
  \caption{\small \label{tab:MCEventFormat}Methods of the \texttt{MCEventFormat} class.}
\end{table*}
\renewcommand{\arraystretch}{1}%

\renewcommand{\arraystretch}{1.5}%
\begin{table*}
  \centering
  \begin{tabular*}{\textwidth}{@{\extracolsep{\fill}}p{.12\textwidth}p{.83\textwidth}@{}} \hline\hline
  \texttt{ctau()}       & Returns, as a floating-point number, the lifetime of the particle
    in millimeters.\\
  \texttt{daughters()}  & Returns, as a vector of pointers to \texttt{MCParticleFormat} objects, a list
    with the daughter particles that are either produced from the decay of the considered particle
    or from its scattering with another particle.\\
  \texttt{momentum()}   & Returns, as a (\textsc{Root}) \texttt{TLorentzVector} object~\cite{Brun:1997pa},
    the four-momentum of the particle. All the properties of the four-momentum can be accessed
    either from the methods associated with the \texttt{TLorentzVector} class, or
    as direct methods of the \texttt{MCParticleFormat} class, after changing the method
    name to be entirely lower case.
    For instance, \texttt{pt()} is equivalent
    to \texttt{momentum().Pt()}. In addition, the methods \texttt{dphi\textunderscore 0\textunderscore 2pi(...)}
    and  \texttt{dphi\textunderscore 0\textunderscore pi(...)} return the difference in azimuthal
    angle normalized in the $[0,2\pi]$ and $[0,\pi]$ ranges, respectively, between the particle
    and any other particle passed as an argument, whereas \texttt{dr(...)} returns their angular distance, the second
    particle being provided as an argument as well.\\
  \texttt{mothers()}    & Returns, as a vector of pointers to \texttt{MCParticleFormat} objects, a list with all the mother
    particles of the considered particle. In the case of an initial particle, this list is empty, while for a decay and a
    scattering process, it contains one and two elements, respectively.\\
   \texttt{mt\textunderscore met()} & Returns, as a floating-point number, the transverse
     mass obtained from a system comprised of the considered particle and the invisible
     transverse momentum of the event. The particles relevant for the
                           calculation must be properly
                           tagged as invisible (see Section~\ref{sec:phys_services}).\\
  \texttt{pdgid()}      & Returns, as an integer number, the Particle Data Group identifier defining the
    nature of the particle. The numbering scheme is based on the Particle Data Group
    review~\cite{Beringer:1900zz}. \\
  \texttt{spin()}       & Returns, as a floating-point number, the cosine of the angle
    between the three-momentum of the particle and its spin vector. This quantity is computed
    in the laboratory reference frame.\\
  \texttt{statuscode()} & Returns, as an integer number, an identifier fixing the initial-,
    intermediate- or final-state nature of the particle. The numbering scheme is based on
    Ref.~\cite{Boos:2001cv}.\\
  \texttt{toRestFrame(...)} & Boosts the four-momentum of the particle
      to the rest frame of a second particle (an \texttt{MCParticleFormat} object given as argument).
      The method modifies the momentum of the particle.\\
  \hline\hline
  \end{tabular*}
  \caption{\small \label{tab:MCParticleFormat}Methods of the \texttt{MCParticleFormat} class.}
\end{table*}
\renewcommand{\arraystretch}{1}%

For instance, assuming that an event sample containing
$N = 10000$ events ($N$ being stored as a double-precision number
in the \texttt{nev} variable) is analyzed, the weight of each event
could be calculated (and stored in the \texttt{wgt} variable
for further use within the analysis) by means of the code sample
\begin{verbatim}
 double lumi = 20000.;
 double nev  = 10000.;
 double wgt = MySample.mc()->xsection()*lumi/nev;
\end{verbatim}
The \texttt{MySample} object is an instance of the
\texttt{SampleFormat} class associated with the sample
being analyzed and we impose the results to be
normalized to 20~fb$^{-1}$ of simulated collisions
(stored in pb$^{-1}$ in the \texttt{lumi} variable).
For efficiency purposes, such a computation should be performed once
and for all at the time of the initialization of the
analysis, and not each time an event is analyzed.
The variable \texttt{wgt} is then promoted
to a member of the analysis class being implemented.

\subsubsection{Internal data format for event handling}\label{sec:dataformat}
\renewcommand{\arraystretch}{1.45}%
\begin{table*}
  \centering
  \begin{tabular*}{\textwidth}{@{\extracolsep{\fill}}p{.14\textwidth}p{.81\textwidth}@{}} \hline\hline
    \texttt{electrons()} & Returns, as a vector of \texttt{RecLeptonFormat} objects,
                           all the reconstructed electrons of the event.\\
    \texttt{jets()}      & Returns, as a vector of \texttt{RecJetFormat} objects,
                           all the reconstructed jets of the event.\\
    \texttt{muons()}     & Returns, as a vector of \texttt{RecLeptonFormat} objects,
                           all the reconstructed muons of the event.\\
    \texttt{photons()}   & Returns, as a vector of \texttt{RecPhotonFormat} objects,
                           all the reconstructed photons of the event.\\
    \texttt{taus()}      & Returns, as a vector of \texttt{RecTauFormat} objects,
                           all the reconstructed hadronic taus of the event.\\
    \texttt{tracks()}    & Returns, as a vector of \texttt{RecTrackFormat} objects,
                           all the reconstructed tracks of the event.\\ \hline
    \texttt{genjets()}   & Returns, as a vector of \texttt{RecJetFormat} objects,
                           all the parton-level jets of the event.\\
    \texttt{MCBquarks()}   & Returns, as a vector of pointers to \texttt{MCParticleFormat} objects,
                           all the parton-level $b$-quarks of the event.\\
    \texttt{MCCquarks()}   & Returns, as a vector of pointers to \texttt{MCParticleFormat} objects,
                           all the parton-level $c$-quarks of the event.\\
    \texttt{MCElectronicTaus()} & Returns, as a vector of pointers to \texttt{MCParticleFormat} objects,
                           all the parton-level
                           tau leptons that have decayed into an electron and a pair of neutrinos.\\
    \texttt{MCHadronicTaus()} & Returns, as a vector of pointers to \texttt{MCParticleFormat} objects,
                           all the parton-level tau leptons that have decayed hadronically.\\
    \texttt{MCMuonicTaus()} & Returns, as a vector of pointers to \texttt{MCParticleFormat} objects,
                            all the parton-level tau leptons that have decayed into a muon and a pair of neutrinos.\\
\hline
    \texttt{MET()}       & Returns, as a \texttt{RecParticleFormat} object,
                           the missing transverse momentum of the event as stored in the event file.\\
    \texttt{MHT()}       & Returns, as a \texttt{RecParticleFormat} object, the missing
                           transverse hadronic momentum $\slashed{\vec{H}}_T$ of the event.\\
    \texttt{TET()}       & Returns, as a floating-point number, the total visible transverse energy of
                           the event $E_T$.\\
    \texttt{THT()}       & Returns, as a floating-point number, the total visible transverse hadronic energy of
                           the event $H_T$.\\
  \hline\hline
  \end{tabular*}
  \caption{\small \label{tab:RecEventFormat}Methods of the \texttt{RecEventFormat} class.}
\end{table*}
\renewcommand{\arraystretch}{1}%

In the \sampleanalyzer\ framework, both Monte Carlo and reconstructed
events are internally handled as instances of a class named
\texttt{EventFormat}. At the time of execution of the analysis
on a specific event, the \texttt{Execute} function receives such an
\texttt{EventFormat} object as its second argument. The properties
of this object reflect those of the current event and can
be retrieved via the two methods
\begin{verbatim}
       event.mc()       event.rec()
\end{verbatim}
which return a pointer to an
\texttt{MCEventFormat} object encompassing
information at the Monte Carlo event level, and a pointer to
a \texttt{RecEventFormat} object specific for managing
information at the reconstructed event level, respectively.

Focusing first on Monte Carlo events, the properties of all
initial-state, in\-ter\-me\-di\-a\-te-state and
final-state particles can be retrieved by means of the \texttt{MCEventFormat} class
(see Table~\ref{tab:MCEventFormat}).
Particles are encoded as instances of the
\texttt{MCParticleFormat} class whose associated methods are
shown in Table~\ref{tab:MCParticleFormat}.
Additionally, general event information,
such as the values for the gauge couplings or the factorization
scale used, is also available if properly stored in the event file.
Finally, the \texttt{MCEventFormat} class also contains
specific methods for the computation of four global event observables:
the amount of (missing) transverse energy $E_T$ ($\slashed{E}_T$)
and of (missing) transverse hadronic energy $H_T$ ($\slashed{H}_T$). 
These quantities are calculated from the transverse momentum
of the final-state particles according to
\be\bsp
  E_T = \sum_{\text{visible particles}} \big| \vec p_T \big|
    \ ,\qquad 
  H_T = \sum_{\text{hadronic particles}} \big| \vec p_T \big|
    \ , \\
  \slashed{E}_T = \big|\slashed{\vec{p}}_T\big| = \bigg| -\sum_{\text{visible particles}} \vec p_T \bigg|
     \ ,\hspace{1.75cm} \\
  \slashed{H}_T = \big|\slashed{\vec{H}}_T\big| = \bigg| -\sum_{\text{hadronic particles}} \vec p_T
    \bigg| \ ,\hspace{1.50cm}
\esp\label{eq:mcmettet}\ee
once the user has defined, in the initialization part
of the analysis, which particles are invisible and which ones are
hadronizing (by means of the configuration functions
described in Section~\ref{sec:phys_services}).
However, the definitions of
Eq.~\eqref{eq:mcmettet} may be not appropriate
if the user wants to
include only specific visible/ha\-dro\-nic particles in the sums.
In this case, he/she should perform their implementation within the
\texttt{Execute} function of the analysis according to his/her needs.
The entire set of properties that can be employed
to analyze a Monte Carlo event is shown in Table~\ref{tab:MCEventFormat}.

For example, the selection of
all the final-state electrons and positrons that are
present in an event and whose transverse momentum
is larger than 50 GeV could be implemented as
\begin{verbatim}
 std::vector<const MCParticleFormat*> electrons;

 for(unsigned int i=0;
     i<event.mc()->particles().size(); i++)
 {
   const MCParticleFormat* prt =
     &event.mc()->particles()[i];

   if(prt->statuscode() != 1) continue;

   if(std::abs(prt->pdgid()) == 11)
   {
     if(prt->momentum().Pt()>50)
       electrons.push_back(prt);
   }
 }
\end{verbatim}
The first line of the code above indicates the declaration of
a vector, dubbed \texttt{electrons}, of pointers to (constant)
\texttt{MC\-Par\-ti\-cle\-For\-mat} objects that contain the selected electrons.
With the next block of \cpp\ commands, we
loop over all the event particles (the \texttt{for} loop)
and store the current particle into a temporary variable \texttt{prt}.
We then discard non-final-state particles,
which have a status code different from one (the first \texttt{if} statement).
Finally, we fill the \texttt{electrons} vector with all electrons and positrons
(with a Particle Data Group
code equal to $\pm 11$, as shown in the second \texttt{if} statement)
whose transverse momentum is greater than 50~GeV (the third \texttt{if}
statement).

We next present the methods that have been designed
for the analysis of reconstructed events and which are
part of the \texttt{RecEventFormat} class. This class
contains functions (see Table~\ref{tab:RecEventFormat})
allowing access to two series of containers,
the first ones gathering final state objects of a given nature and
the second ones collecting specific generator-level (or equivalently parton-level)
objects. All these containers can be further employed within an analysis
so that the properties of the different objects can be retrieved and
subsequently used, \textit{e.g.}, for cuts and histograms.
All the available methods associated with reconstructed objects
have been collected in Table~\ref{tab:RecParticleFormat}, while we recall
that the \texttt{MCParticleFormat} class has been described in
Table~\ref{tab:MCParticleFormat} (necessary
for the handling of generator-level objects).
In the case where some pieces of information
(either specific properties of a given particle species or a given container
itself) are absent from the event file, the related methods return
null results.

\renewcommand{\arraystretch}{1.5}%
\begin{table*}
  \centering
  \begin{tabular*}{\textwidth}{@{\extracolsep{\fill}}p{.11\textwidth}p{.84\textwidth}@{}} \hline\hline
    \texttt{btag()}      & This method is specific to \texttt{RecJetFormat} objects and returns a boolean quantity
                           describing whether the jet has been tagged as a $b$-jet.\\
    \texttt{ctag()}      & This method is specific to \texttt{RecJetFormat} objects and returns a boolean quantity
                           describing whether the jet has been tagged as a $c$-jet.\\
    \texttt{charge()}    & Returns, as an integer number, the electric charge of the object
                           (relative to the fundamental unit of electric charge $e$).
                           This method is available
                           for the \texttt{RecLeptonFormat}, \texttt{RecTauFormat} and \texttt{RecTrackFormat} classes.\\
    \texttt{etaCalo()}   & This method is specific to the \texttt{RecTrackFormat} class and returns,
                           as a floating-point number,
                           the pseudorapidity corresponding to the entry point of the track in the calorimeter.\\
    \texttt{isolCones()} & Returns a vector of pointers to instances of the \texttt{IsolationConeType} class.
       This class allows one to retrieve information about the isolation of the object
       after defining a cone of a given size
       (a floating-point number accessed via the \texttt{deltaR()} method of the class) centered on it. The (integer)
       number of tracks in the cone is obtained by means of the \texttt{ntracks()} method, the sum of the
       transverse momenta of these tracks by means of the \texttt{sumPT()} method and the amount of
       calorimetric (transverse) energy in the cone
       by means of the \texttt{sumET()} method. The \texttt{isolCones()}
       method has only been implemented for the \texttt{RecTrackFormat}, \texttt{RecLeptonFormat},
       \texttt{RecPhotonFormat} and \texttt{RecJetFormat} classes. A modified version of \textsc{Delphes 3}
       that supports this structure has been introduced in
       Ref.~\cite{Brooijmans:2014eja}.\\
    \texttt{momentum()}  & Returns, as a (\textsc{Root}) \texttt{TLorentzVector} object~\cite{Brun:1997pa},
        the four-momentum of the particle. This method is available for all types of
        reconstructed objects. All the properties of the four-momentum can
        be accessed either from the methods associated with the \texttt{TLorentzVector}
        class, or as direct methods of the different classes of objects, after changing the
        method name to be entirely lower case.
        For instance, the method \texttt{pt()} is equivalent
                           to \texttt{momentum().Pt()}. In addition, the methods
                           \texttt{dphi\textunderscore 0\textunderscore 2pi(...)}
                           and  \texttt{dphi\textunderscore 0\textunderscore pi(...)} return the difference in azimuthal
                           angle normalized in the $[0,2\pi]$ and $[0,\pi]$ ranges, respectively, between the object
                           and any other object passed as an argument, whereas \texttt{dr(...)} returns their
                           angular distance, the second
                           object being provided as an argument as well.\\
    \texttt{mt\textunderscore met()} & Returns, as a floating-point number, the transverse
     mass obtained from a system comprised of the considered particle and the missing
     transverse momentum of the event.\\
    \texttt{ntracks()}   & Returns, as an integer number, the number of charged tracks associated with
                           the reconstructed object. This method has been implemented for the  \texttt{RecTauFormat}
                           and \texttt{RecJetFormat} classes.\\
    \texttt{pdgid()}     & This method is specific to the \texttt{RecTrackFormat} class and returns,
                           as an integer number, the Particle Data Group
                           identifier defining the nature of the particle giving rise to the track. The numbering
                           scheme is based on the Particle Data Group review~\cite{Beringer:1900zz}.\\
    \texttt{phiCalo()}   & This method is specific to the \texttt{RecTrackFormat} class and returns,
                           as a floating-point number,
                           the azimuthal angle with respect to the beam direction corresponding to the entry point
                           of the track in the calorimeter.\\
    \texttt{sumET\textunderscore isol()} & Returns, as a floating-point number, the amount of
       calorimetric (transverse) energy lying in a specific
       cone centered on the object. The cone size is fixed at the level of
       the detector simulation and this method is available for the \texttt{RecLeptonFormat} class
       (this information is available in the \textsc{Lhco} format).\\
    \texttt{sumPT\textunderscore isol()} & Returns, as a floating-point number, the sum of the
       transverse momenta of all tracks lying in a given cone centered on
       the object. The cone size is fixed at the level of
       the detector simulation and this method is available for the \texttt{RecLeptonFormat} class
       (this information is available in the \textsc{Lhco} format).\\
    \texttt{EEoverHE()}  & Returns, as a floating-point number, the ratio of the electromagnetic to hadronic
                           calorimetric energy associated with the object. This method is available
                           for the \texttt{RecLeptonFormat}, \texttt{RecTauFormat} and \texttt{RecJetFormat} classes.\\
    \texttt{ET\textunderscore PT\textunderscore isol()} & Returns, as a floating-point number, the amount
       of calorimetric (transverse) energy lying in a given cone centered on the object
       calculated relatively to the sum
       of the transverse momentum of all tracks in this cone. The cone size is fixed at the level of
       the detector simulation and this method is available for the \texttt{RecLeptonFormat} class
       (this information is available in the \textsc{Lhco} format).\\
    \texttt{HEoverEE()}  & Returns, as a floating-point number, the ratio of the hadronic to electromagnetic
                           calorimetric energy associated with the object. This method is available
                           for the \texttt{RecLeptonFormat}, \texttt{RecTauFormat} and \texttt{RecJetFormat}
                           classes.\\
  \hline\hline
  \end{tabular*}
  \caption{\small \label{tab:RecParticleFormat}Methods giving access the properties of
    the reconstructed objects represented as instances of the \texttt{RecLeptonFormat},
    \texttt{RecJetFormat}, \texttt{RecPhotonFormat}, \texttt{RecTauFormat}, \texttt{RecTrackFormat}
    and \texttt{RecParticleFormat} classes.}
\end{table*}
\renewcommand{\arraystretch}{1}%

Finally, as for the \texttt{MCEventFormat} class, specific functions (see
Table~\ref{tab:RecEventFormat}) have been implemented
to access the (missing) transverse energy and (missing) hadronic transverse
energy of the event. While the value of the $\slashed{E}_T$ variable is taken
from the event file and not calculated on the fly, the other variables
are computed from the information on the reconstructed objects,
\be\bsp
  E_T =&\ \sum_{\text{jets, charged leptons, photons}} \big| \vec p_T \big|
    \ , \\
  H_T =&\ \sum_{\text{jets}} \big| \vec p_T \big|
    \ , \\
  \slashed{H}_T = &\ \big|\slashed{\vec{H}}_T\big| = \bigg| -\sum_{\text{jets}} \vec p_T
    \bigg| \ .
\esp\label{eq:recglobal}\ee

As an example, we show how
an isolation requirement on final-state muons can be implemented.
To do this we define an isolation variable $I_{\rm rel}$ as
the amount of transverse energy,
relative to the transverse momentum of the muon,
present in a cone of radius $R = 0.4$ centered on the muon.
We constrain this quantity to satisfy $I_{\rm rel} < 20\%$. A possible
corresponding sample of \cpp\ code is
\begin{verbatim}
 std::vector<const RecLeptonFormat*> MyMuons;
 for(unsigned int i=0;
     i<event.rec()->muons().size(); i++)
 {
   const RecLeptonFormat *Muon =
     &event.rec()->muons()[i];

   for(unsigned int j=0;
       j<Muon->isolCones().size(); j++)
   {
     const IsolationConeType *cone = 
       &Muon->isolCones()[j];

     if(std::fabs(cone->deltaR()-0.4)<1e-3)
     {
       if(cone->sumET()/Muon->momentum().Pt()<.20)
         MyMuons.push_back(Muon);
     }
   }
 }
\end{verbatim}
With those lines of code, we start by declaring the \texttt{MyMuons}
variable, a vector of pointers to
\texttt{RecLeptonFormat} objects, that will refer
to the reconstructed muons tagged as isolated. Then, we proceed with
a \texttt{for}-loop dedicated to the computation of
the $I_{\rm rel}$ variable for each of the final state muons. In the case where
$I_{\rm rel}$ is smaller than 20\%, the muon is added to the \texttt{MyMuons}
container. In more detail, this \texttt{for}-loop works as follows. The current muon is
stored in a temporary variable called \texttt{Muon}. The calculation
of $I_{\rm rel}$ relies, first, on the amount of calorimetric energy in
a cone of radius \mbox{$R=0.4$} centered on the muon and second,
on the transverse momentum of the current muon.
The first of these two quantities is evaluated via the \texttt{isolCones()} method
of the \texttt{RecLeptonFormat} class (see Table~\ref{tab:RecParticleFormat})
whereas the second one is derived from the
muon four-momentum (obtained from the \texttt{momentum()} me\-thod of the
\texttt{RecLeptonFormat} class). In the example above,
we assume that information on muon isolation associated with several cone sizes
is available, including the choice \mbox{$R=0.4$}. The second
\texttt{for}-loop that has been implemented selects the desired
value of $R$. The subsequent computation of the $I_{\rm rel}$ quantity
is immediate.
We refer to Ref.~\cite{Brooijmans:2014eja} for more detailed
examples on this topic, in cases where event simulation is
based on a modified version of \textsc{Delphes 3} properly handling
such a structure for the isolation information.

\renewcommand{\arraystretch}{1.5}%
\begin{table*}
  \centering
  \begin{tabular*}{\textwidth}{@{\extracolsep{\fill}}p{.10\textwidth}p{.85\textwidth}@{}} \hline\hline
    \texttt{DisableColor()} & Switches off the display of messages in color.
                              Colors are switched on by default, and the color scheme is hard-coded.\\
    \texttt{EnableColor()}  & Switches on the display of messages in color.
                              Colors are switched on by default, and the color scheme is hard-coded.\\
    \texttt{Mute()}         & Switches off a specific message service. Services are switched on by default.\\
    \texttt{SetStream(...)}  & Takes a pointer of type \texttt{ofstream} as an
      argument and redirect the output of a given service to a file.\\
    \texttt{UnMute()}       & Switches on a specific message service. Services are switched on by default.\\
\hline\hline
  \end{tabular*}
  \caption{\small \label{tab:msg}Methods associated with a given message service. The available services are
   \texttt{INFO}, \texttt{WARNING}, \texttt{ERROR} and \texttt{DEBUG}.}
\end{table*}
\renewcommand{\arraystretch}{1}%

\subsubsection{Applying cuts and filling histograms}\label{sec:cutshistos}
The cuts for the analysis, having been declared in the \texttt{I\-ni\-ti\-a\-li\-ze}
function (see Section~\ref{sec:init}),
are applied in the \texttt{Execute} function by means
of the \texttt{RegionSelectionManager} method \texttt{ApplyCut}
(see Table~\ref{tab:RSMfcts}). Its two arguments consist
of a boolean quantity governing the cut condition (\textit{i.e.}, it indicates
whe\-ther the current event satisfies this cut) and a string which should be the name of one
of the declared cuts.

This method starts by cycling through all regions
associated with this cut. For each region, it checks whether the region is still
surviving all cuts applied so far by evaluating an internal boolean variable.
If a given region is found to be already failing one of the preceding
cuts (indicated by the internal surviving variable having the value
\texttt{false}), the \texttt{ApplyCut}
method continues with the next region associated with the considered cut.
On the other hand if the region is surviving, the cut-flow
for this region is updated according to
the cut condition (the boolean argument of the
\texttt{ApplyCut} method) and the internal
surviving variable will be kept as \texttt{true} or changed to \texttt{false}
as appropriate.
The aforementioned internal boolean variables indicating the survival of each region
should all be initialized to \texttt{true} 
when starting to analyze a given event.
This is achieved by adding,
at the beginning of the \texttt{Execute} function,
\begin{verbatim}
 Manager()->InitializeForNewEvent(myWeight);
\end{verbatim}
where \texttt{MyWeight} is a floating-point number representing the weight of the event.
The weight is used when histograms are filled and cut-flow charts
calculated, and can be modified within the analysis by making use
of the \texttt{Set\-Cur\-rent\-E\-vent\-Weight}
method of the \texttt{RegionSelectionManager} class.

The analysis manager also stores internally the total number of surviving regions,
which is updated when a specific region fails a cut. This enables the \texttt{ApplyCut}
method to determine and return,
after cycling through the associated \texttt{Re\-gion\-Se\-lec\-ti\-on} instances,
a boolean quantity which is set to \texttt{false} in the case where not a single surviving region
remains. The output of the \texttt{ApplyCut} method
is thus equal to the boolean value of the statement {\it there is at least one region in the analysis,
not necessarily one of those associated with this specific cut, which is still passing all cuts so far}.
When it switches from \texttt{true} to \texttt{false},
the present event should no longer be analyzed, and one should move on with the next event.
It is therefore recommended, for efficiency purposes,
to always call the \texttt{ApplyCut} method in the following schematic manner,
\begin{verbatim}
 if ( !ApplyCut(...) )
   return true;
\end{verbatim}
with the \texttt{return} command terminating the analysis of the current event
if all regions are failing the set of cuts applied so far.

Since, trivially, cuts keep some events and reject others, the distribution of an
observable is affected by the placement of its histogram-filling command
within the full sequence of cuts.
Then since each region has its own unique set of cuts (by definition), the
distribution of any observable is in general different for any two regions.
However, it is meaningful to consider a single histogram as associated with multiple
regions, {\it if} it is filled before any cuts are made that distinguish the regions.
As an example, a typical format for processing an event would be a set of common
preselection cuts, then the filling of various histograms (which are thus
associated with all regions), then the application of the region-specific cuts
(possibly followed by some further histogramming).

In \madanalysis, we deal with this within
the histogram-filling method of the \texttt{RegionSelectionManager}
class, \texttt{Fill\-His\-to}, which takes as arguments a string and
a floating-point number. The string
should be the name of one of the declared histograms, and the floating-point
number represents the value of the histogrammed observable for the event under
consideration. This method can be called as in
\begin{verbatim}
 Manager()->FillHisto("ptl1", val);
\end{verbatim}
where \texttt{"ptl1"} is the name of the considered histogram
(continuing with the example from Section~\ref{sec:init})
and \texttt{val} is the value of the observable of interest,
namely the transverse momentum of the leading lepton in our case.
The \texttt{FillHisto} method begins by
verifying whether each of the regions associated with this histogram
is surviving all cuts applied so far (via the internal surviving variable
above-mentioned). In the case where all the associated regions are found
surviving (failing) the cuts, the histogram is (not) filled.
If a mixture of surviving and non-surviving regions is found,
the program stops and displays an error message to the screen, as
this situation implies that the histogram filling command has been called
{\it after} at least one cut yields a distinction among the associated
regions. This indicates an error in the design of the analysis.

\subsubsection{Finalizing an analysis}
Once all the events have been processed, the program calls
the function \texttt{Finalize}. The user can make use
of it for drawing histograms or
deriving cut-flow charts as indicated in the manual for older versions of the
program~\cite{Conte:2012fm}; however, from the version
of \madanalysis\ introduced in this paper onwards,
the \texttt{Finalize} function does not need to be implemented
anymore. Output files written according to
the \textsc{Saf} format (see Section~\ref{sec:saf}) are automatically generated.

\subsubsection{Message services}
\label{sec:msg_services}

The \cpp\ core of \madanalysis\ includes a class of functions
dedicated to the display of text on the screen at the time of the
execution of the analysis.
Whereas only two distinct levels of message are accessible
by using the standard \cpp\ streamers
(\texttt{std::cout} and \texttt{std:cerr} for normal
and error messages), the \sampleanalyzer\ library enables the user
to print messages that can be classified into four categories.
In this way,
information (the \texttt{INFO} function),
warning (the \texttt{WARNING} function),
error (the \texttt{ERROR} function) and
debugging (the \texttt{DEBUG} function)
messages can be displayed as in the following sample of code,
\begin{verbatim}
 INFO    << "..." << endmsg;
 WARNING << "..." << endmsg;
 ERROR   << "..." << endmsg;
 DEBUG   << "..." << endmsg;
\end{verbatim}
Additionally, warning and error messages provide information on the
line number of the analysis code that is at the source of the message.
The effect of a given message service can finally be modified by
means of the methods presented in Table~\ref{tab:msg}.

\subsubsection{Physics services}
\label{sec:phys_services}

\renewcommand{\arraystretch}{1.5}%
\begin{table*}
  \centering
  \begin{tabular*}{\textwidth}{@{\extracolsep{\fill}}p{.255\textwidth}p{.665\textwidth}@{}} \hline\hline
    \texttt{mcConfig().AddHadronicId(...)} & Adds a particle species, identified
       via its Particle Data Group code (an integer number given as argument), to the list of
       hadronizing particles. Mandatory for the computation of $H_T$
       and $\slashed{H}_T$ in the case of Monte Carlo events (see Section~\ref{sec:dataformat}).\\
    \texttt{mcConfig().AddInvisibleId(...)} & Adds a particle species, identified
       via its Particle Data Group code (an integer number given as argument),
       to the list of invisible particles. Mandatory for the computation of $E_T$
       and $\slashed{E}_T$ in the case of Monte Carlo events
       (see Section~\ref{sec:dataformat}).\\
    \texttt{mcConfig().Reset()} & Reinitializes the lists of invisible and hadronizing particles to empty lists.\\
    \texttt{recConfig().Reset()} & Defines (reconstructed) leptons as isolated when no jet is present in a cone
       of radius $R = 0.5$ centered on the lepton.\\
    \texttt{recConfig().UseDeltaRIsolation(...)} & Defines (reconstructed) leptons as isolated when no jet is present
       in a cone, with a radius given as a floating-point number in argument, centered on the lepton.\\
    \texttt{recConfig().UseSumPTIsolation(...)} & Defines (reconstructed) leptons as isolated when both
       the sum $\Sigma_1$ of the transverse momenta of all tracks in a cone
       (of radius fixed at the level of the detector simulation)
       centered on the lepton is smaller than a specific threshold (the first argument) and
       the amount of calorimetric energy in this cone, relative to $\Sigma_1$,
       is smaller than another threshold (the second argument).
       This uses the information provided
       by the \texttt{sumPT\textunderscore isol()}
       and \texttt{ET\textunderscore PT\textunderscore isol()} methods
       of the \texttt{RecLeptonFormat} class (see Table~\ref{tab:RecParticleFormat}).\\
    \texttt{Id->IsBHadron(...)} & Returns a boolean quantity indicating whether an \texttt{MCParticleFormat}
      object passed as argument is a hadron originating from the fragmentation of a $b$-quark.\\
    \texttt{Id->IsCHadron(...)} & Returns a boolean quantity indicating whether an \texttt{MCParticleFormat}
      object passed as argument is a hadron originating from the fragmentation of a $c$-quark.\\
    \texttt{Id->IsFinalState(...)} & Returns a boolean quantity indicating whether an \texttt{MCParticleFormat} object
      passed as argument is one of the final-state particles of the considered event.\\
    \texttt{Id->IsHadronic(...)} & Returns a boolean quantity indicating whether an \texttt{MCParticleFormat}
      or a reconstructed object passed as argument yields any hadronic activity in the event.\\
    \texttt{Id->IsInitialState(...)} & Returns a boolean quantity indicating whether an \texttt{MCParticleFormat} object
      passed as argument is one of the initial-state particles of the considered event.\\
    \texttt{Id->IsInterState(...)} & Returns a boolean quantity indicating whether an \texttt{MCParticleFormat} object
      passed as argument is one of the intermediate-state particles of the considered event.\\
    \texttt{Id->IsInvisible(...)} & Returns a boolean quantity indicating whether an \texttt{MCParticleFormat}
      or a reconstructed object passed as argument gives rise to missing energy.\\
    \texttt{Id->IsIsolatedMuon(...)} & Returns a boolean quantity indicating whether a \texttt{RecLeptonFormat}
      object passed as a first argument is isolated within a given reconstructed event, passed as a
      second argument (under the format of a \texttt{RecEventFormat} object).\\
    \texttt{Id->SetFinalState(...)} & Takes an \texttt{MCEventFormat} object as argument and defines
      the status code number associated with final-state particles.\\
    \texttt{Id->SetInitialState(...)} & Takes an \texttt{MCEventFormat} object as argument and defines
      the status code number associated with initial-state particles.\\
    \texttt{Transverse->AlphaT(...)} & Returns the value of the $\alpha_T$ variable~\cite{Randall:2008rw},
      as a floating-point number,
      for a given (Monte Carlo or reconstructed) event passed as argument.\\
    \texttt{Transverse->MT2(...)} & Returns, as a floating-point number, the value of the $M_{T2}$
      variable~\cite{Lester:1999tx,Cheng:2008hk} computed from a system of two visible objects (the first two arguments, any particle
      class being accepted),
      the missing momentum (the third argument) and a test mass (a floating-point number given as the last argument).\\
    \texttt{Transverse->MT2W(...)} & Returns, as a floating-point number, the value of the $M_{T2}^W$
      variable~\cite{Bai:2012gs} computed from a system of jets (a vector of \texttt{RecJetFormat}
      objects in the first argument), a visible particle (given as the second argument, any particle class being accepted) and
      the missing momentum (the third argument). Only available for reconstructed events.\\
\hline\hline
  \end{tabular*}
  \caption{\small \label{tab:physics}Physics service methods.}
\end{table*}
\renewcommand{\arraystretch}{1}%

The \sampleanalyzer\ core includes a series of built-in functions aiming to facilitate
the writing of an analysis from the user viewpoint. More precisely, these
functions are specific for
particle identification or observable calculation and have
been grouped into several subcategories of the \cpp\ pointer
\texttt{PHY\-SICS}. All the available methods are listed in Table~\ref{tab:physics},
and we provide, in the rest of this section, a few more details,
together with some illustrative examples.

As mentioned in Section~\ref{sec:dataformat}, \madanalysis\ can
compute the (missing) transverse energy and (missing) hadronic transverse
energy associated with a given Monte Carlo event. This calculation however
relies on a correct identification of the invisible and hadronizing particles.
This information must be provided
by means of the \texttt{mcConfig()} category of physics services, as
for instance, in
\begin{verbatim}
 PHYSICS->mcConfig().AddInvisibleId(1000039);
 PHYSICS->mcConfig().AddHadronicId(5);
\end{verbatim}
These intuitive lines of code indicate to the program that the gravitino
(whose Particle Data Group identifier is 1000039)
yields missing energy and that the bottom quark (whose Particle Data
Group identifier is 5) will eventually hadronize.

An important category of methods shipped with the physics services
consists of functions dedicated to the identification of particles
and to the probing of their nature (invisible, hadronizing,
\textit{etc.}). They are collected within the \texttt{Id} structure
attached to the \texttt{PHYSICS} object. For instance (see Table~\ref{tab:physics}
for the other methods),
\begin{verbatim}
 PHYSICS->Id->IsInvisible(prt)
\end{verbatim}
allows one to test the (in)visible nature
of the particle referred to by the pointer \texttt{prt}. Also, basic
isolation tests on \texttt{RecLeptonFormat} objects can be performed when analyzing
reconstructed events. Including in the analysis
\begin{verbatim}
 PHYSICS->Id->IsIsolatedMuon(muon, event)
\end{verbatim}
yields a boolean value related to the (non-)isolated
nature of the reconstructed lepton \texttt{muon}, \texttt{event} being here a
\texttt{Rec\-E\-vent\-For\-mat} object.
Two isolation algorithms can be employed.
By default, the program verifies that no reconstructed jet lies in a
cone of radius $R=0.5$ centered on the lepton. The value of
$R$ can be modified via the \texttt{recConfig()} category of physics
services,
\begin{verbatim}
 PHYSICS->recConfig().UseDeltaRIsolation(dR);
\end{verbatim}
where \texttt{dR} is a floating-point variable with the chosen cone size.
The user can instead require the program to tag leptons as isolated when
both the sum of the transverse momenta of all tracks in a
cone (of radius fixed at the level of the detector simulation)
centered on the lepton is smaller than a specific threshold
and when the amount of calorimetric energy in this cone, calculated
relative to the sum of the transverse momenta of all tracks in the cone,
is smaller than another threshold. This uses the information provided
by the \texttt{sumPT\textunderscore isol()}
and \texttt{ET\textunderscore PT\textunderscore isol()} methods
of the \texttt{RecLeptonFormat} class (see Table~\ref{tab:RecParticleFormat}) and
can be activated by implementing
\begin{verbatim}
 PHYSICS->recConfig().UseSumPTIsolation(sumpt,et_pt);
\end{verbatim}
where \texttt{sumpt} and \texttt{et\textunderscore pt} are the two mentioned
thresholds.
For more sophisticated isolation tests,
such as those based on the information encompassed
in \texttt{IsolationConeType} objects possibly provided for reconstructed
jets, leptons and photons (see Section~\ref{sec:dataformat}), it is left to the user
to manually implement the corresponding routines in his/her analysis.

In addition to identification routines, physics services include
built-in functions allowing one to compute global event
observables, such as several
transverse variables that are accessible through the
\texttt{Transverse} structure attached to the \texttt{PHY\-SICS} object.
More information on the usage of these methods is provided in Table~\ref{tab:physics}.

\subsubsection{Sorting particles and objects}
In most analyses, particles of a given species are identified
according to an ordering in their transverse momentum or energy.
In contrast, vector of particles as returned after the reading
of an event are in general unordered and therefore need to be sorted.
This can be achieved by means of sorting routines that can be called
following the schematic form:
\begin{verbatim}
  SORTER->sort(parts, crit)
\end{verbatim}
In this line of code, \texttt{parts} is a vector of
(Monte Carlo or reconstructed) objects and \texttt{crit}
consists of the ordering criterion. The allowed choices
for the latter
are \texttt{ETAordering} (ordering in pseudorapidity),
\texttt{ETordering} (ordering in transverse energy),
\texttt{Eordering} (ordering in energy), \texttt{Pordering}
(ordering in the norm of the three-momentum),
\texttt{PTordering} (ordering in the transverse momentum), \texttt{PXordering}
(ordering in the $x$-component of the three-momentum), \texttt{PYordering}
(ordering in the $y$-component of the three-momentum) and
\texttt{PZordering} (ordering in the $z$-component of the three-momentum).
The objects are always sorted in terms of decreasing values of the
considered observable.

\subsection{Compiling and executing the analysis}
\label{sec:exec}
In Section~\ref{sec:template}, we have pointed out that the \texttt{Build}
subdirectory of the analysis template contains
a \texttt{Makefile} script readily to be used.
In this way, the only task left to the user after having implemented his/her analysis
is to launch this script in a shell, directly from the \texttt{Build} directory.
This leads first to the creation of a library that is stored in the \texttt{Build/Lib}
subdirectory, which includes all the analyses implemented
by the user and the set of classes and methods of the \sampleanalyzer\
kernel. Next, this library is linked to the main program and
an executable named \texttt{MadAnalysis5Job} is generated (and stored in the
\texttt{Build} directory).

The program can be run by issuing in a shell the command
\begin{verbatim}
 ./MadAnalysis5Job <inputfile>
\end{verbatim}
where \texttt{<inputfile>} is a text file with a list of paths to
all event files to analyze. All implemented analyses are sequentially
executed and the results, generated according to the \textsc{Saf}
format (see Section~\ref{sec:saf}), are stored in the \texttt{Output}
directory.

\subsection{The structure of the output of an analysis}
\label{sec:saf}
As indicated in the previous section, the program stores, after its execution,
the results of the analysis or analyses that have
been implemented by the user in the \texttt{Output} subdirectory of
the working directory. First, a subdirectory with the same name as the input file
(\texttt{<inputfile>} in the schematic example of Section~\ref{sec:exec}) is created.
If a directory with this name exists already, the code uses
it without deleting its content. It contains a \textsc{Saf} file (updated if
already existing) with global information
on the analyzed event samples organized following an \textsc{Xml}-like syntax:
\begin{verbatim}
 <SampleGlobalInfo>
  # xsection  xsec_error  nevents  sum_wgt+  sum_wgt-
  0.00e+00    0.00e+00    0        0.00e+00  0.00e+00
 </SampleGlobalInfo>
\end{verbatim}
where we have set the numerical values to zero for the sake of the illustration.
In reality these values are extracted from the event file that is read;
they are kept equal to zero
if not available. In addition, the format includes header and footer tags
(\texttt{SAFheader} and \texttt{SAFfooter}) omitted for brevity.

Secondly, a subdirectory specific to each of the executed analyses is created within
the \texttt{<inputfile>} directory. The name of the subdirectory
is the name of the associated analysis followed by an integer number chosen in such
a way that the directory name is unique. This directory contains a \textsc{Saf}
file with general information
on the analysis (\texttt{name.saf}, \texttt{name} denoting
a generic analysis name), a directory with histograms
(\texttt{Histograms}) and a directory with cut-flow charts (\texttt{Cutflows}).

In addition to a header and a footer, the \texttt{name.saf} file, still encoded
according to an \textsc{Xml}-like structure, contains a list
with the names of the regions that have been declared in the analysis implementation.
They are embedded in a \texttt{Re\-gi\-on\-Se\-lec\-ti\-on} \textsc{Xml} structure, as in
\begin{verbatim}
 <RegionSelection>
  "MET>200"
  "MET>300"
 </RegionSelection>
\end{verbatim}
when taking the example of Section~\ref{sec:init}.

The \texttt{Histograms} subdirectory contains a unique \textsc{Saf} file
with, again in addition to a possible header and footer, all the histograms
implemented by the user. The single histogram declared in Section~\ref{sec:init} would
be encoded in the \textsc{Saf} format as in the following self-explanatory lines of code:
\begin{verbatim}
 <Histo>
  <Description>
   "ptl1"
   # nbins        xmin           xmax
   20             50             500
   # associated RegionSelections
   MET>300   # Region nr. 1
  </Description>
  <Statistics>
   0 0 # nevents
   0 0 # sum of event-weights over events
   0 0 # nentries
   0 0 # sum of event-weights over entries
   0 0 # sum weights^2
   0 0 # sum value*weight
   0 0 # sum value^2*weight
   0 0 # sum value*weight^2
 </Statistics>
 <Data>
  0 0 # number of nan
  0 0 # number of inf
  0 0 # underflow
  0 0 # bin 1 / 20
  ...
  0 0 # bin 20 / 20
  0 0 # overflow
  </Data>
</Histo>
\end{verbatim}
where the dots stand for the other bins that we have omitted for brevity.
Again, for the
sake of the example we have set all values to zero.

Finally, the \texttt{Cutflows} directory contains one
\textsc{Saf} file for each of the declared regions, the filename
being the name of the region followed by the \texttt{saf}
extension. Each of these files contains
the cut-flow chart associated with the considered region encoded
by means of two types of \textsc{Xml} tags. The first one is only
used for the initial number of events (\texttt{I\-ni\-ti\-al\-Coun\-ter})
whereas the second one is dedicated to
each of the applied cuts. Taking
the example of the first of the two cuts declared in Section~\ref{sec:init},
the \texttt{MET\textunderscore gr\textunderscore200.saf} file (the \texttt{>}
symbol in the region name has been replaced by \texttt{\textunderscore gr\textunderscore})
would read
\begin{verbatim}
 <InitialCounter>
  "Initial number of events"    #
  0          0                  # nentries
  0.00e+00   0.00e+00           # sum of weights
  0.00e+00   0.00e+00           # sum of weights^2
 </InitialCounter>
 <Counter>
  "1lepton"                     # 1st cut
  0          0                  # nentries
  0.00e+00   0.00e+00           # sum of weights
  0.00e+00   0.00e+00           # sum of weights^2
 </Counter>
\end{verbatim}
which is again self-explanatory.

\section{Illustrative examples}\label{sec:examples}
In this section, we show two examples of analyses making use of the new features
of \madanalysis\ introduced in the previous section.
First, we focus in Section~\ref{sec:stop} on the reinterpretation of a CMS search
for stops in 8~TeV events with one single lepton, jets
and missing energy~\cite{Chatrchyan:2013xna}. Second,
we investigate in Section~\ref{sec:monotop}
the implementation of a recent phenomenological analysis dedicated fromthe study
of monotop systems decaying in the hadronic mode~\cite{Agram:2013wda}.

\subsection{Recasting a CMS search for supersymmetric partners of the top quark}\label{sec:stop}

We present an implementation of
the CMS cut-based strategy for probing stops in the single lepton and missing energy
channel as presented in Ref.~\cite{Chatrchyan:2013xna}.
The analysis contains 16 overlapping signal regions that share
a set of common preselection cuts and that are distinguished
by extra requirements. We refer
to Refs.~\cite{Chatrchyan:2013xna,Brooijmans:2014eja,Dumont:2014tja}
for more details on the cuts, the analysis
regions and their reimplementation that strictly obeys the syntax introduced
in Section~\ref{sec:ma5}.
For simplicity, we
consider below only one of the signal regions, which is defined by
the following cuts.
\begin{itemize}
 \item We select events with a single tightly-isolated
  electron (mu\-on) with a transverse momentum $p_T^\ell>30$~GeV
  (25~GeV) and a pseudorapidity satisfying $|\eta^\ell|<1.4442$ (2.1). Isolation
  is enforced by constraining the sum of the transverse momenta of all
  tracks\footnote{The original analysis defines isolation from particle-flow
  objects~\cite{PFT-09-001}. Their correct modeling being difficult to reproduce in
  our setup, we only consider tracks in the inner detector.}
  in a cone of radius $R=0.3$ centered on the lepton to be smaller than min(5~GeV,
  $0.15 p_T^\ell$).
 \item Events featuring in addition a loosely-isolated lepton with a transverse
   momentum $p_T^\ell > 5$~GeV are vetoed. Isolation is enforced by constraining
   the sum of the transverse momenta of all tracks in a cone of $R=0.3$ centered
   on the lepton to be smaller than $0.20 p_T^\ell$.
 \item Events featuring an isolated track of transverse momentum
   \mbox{$p_T^{\rm track} > 10$~GeV} and whose electric charge is opposite to the
   one of the primary lepton are vetoed. Isolation is enforced by constraining
   the sum of the transverse momenta of all tracks in a cone of radius $R=0.3$ centered
   on the track to be smaller than $0.10 \, p_T^{\rm track}$.
 \item Events with reconstructed hadronic taus of
   transverse momentum greater than 20~GeV are vetoed.
 \item Jets present in a cone of radius $R=0.4$ centered on a lepton are discarded.
  Four central jets with a transverse momentum $p_T^j > 30$~GeV and a
   pseudorapidity $|\eta^j| < 2.4$ are then required, with at least
   one of them being $b$-tagged.
 \item The transverse mass $M_T$ reconstructed from the lepton
  and the missing transverse momentum is constrained to be larger than 120~GeV.
 \item The signal region consider events with at least 300~GeV of missing
   transverse energy, the preselection thre\-shold common to all the analysis regions
   being 100~GeV. The missing transverse momentum is also required to be separated from
   the two hardest jets ($\Delta \phi > 0.8$, $\phi$ denoting the azimuthal
   angle with respect to the beam direction).
  \item The final-state configuration is required to contain a ha\-dro\-ni\-cal\-ly
   decaying top quark
   (by means of a $\chi^2$-fit based on the reconstructed objects).
  \item The $M_{T2}^W$ variable must be greater than 200~GeV.
\end{itemize}

For the sake of the example,
we include below a snippet of code describing the implementation of the veto
on the presence of isolated tracks.
We start by creating a variable (\texttt{Tracks}) containing all the tracks
whose transverse momentum is larger than 10~GeV and pseudorapidity satisfies
$|\eta^{\rm track}|<2.1$:
\begin{verbatim}
 for(unsigned int i=0;
     i<event.rec()->tracks().size(); i++)
 {
   const RecTrackFormat *myTrack =
     &(event.rec()->tracks()[i]);
   double pt = myTrack->momentum().Pt();
   double abseta = std::fabs(myTrack->eta());
   if(pt>10.&&abseta<2.1) Tracks.push_back(myTrack);
 }
\end{verbatim}
Next, we iterate over this container and update a boolean variable
\texttt{noIsolatedTrack} (initialized to \texttt{true})
to \texttt{false} if
an isolated track of opposite charge (compared to
the primary lepton charge, stored in the \texttt{LeptonCharge} variable)
is found:
\begin{verbatim}
 noIsolatedTrack = true;
 for(unsigned int i=0; i<Tracks.size(); i++)
 {
   if(Tracks[i]->charge()!=LeptonCharge)
   {
     for(unsigned int j=0; 
         j<Tracks[i]->isolCones().size(); j++)
     {
      const IsolationConeType *cone =
         &Tracks[i]->isolCones()[j];
      double pt = Tracks[i]->momentum().Pt();
      if(std::fabs(cone->deltaR()-0.3)<0.001)
        if(cone->sumPT()<.1*pt)
          { noIsolatedTrack = false; break; }
     }
   }
 }
\end{verbatim}
It is then sufficient to implement the verification of the cut condition as explained
in Section~\ref{sec:cutshistos},
\begin{verbatim}
 if(!Manager()->ApplyCut(NoIsolatedTrack,"VetoIsTr"))
   return;
\end{verbatim}
where we assume that a cut
\texttt{"VetoIsTr"} has been declared in the analysis initialization
method.

The validation of our reimplementation relies on the comparison
of results obtained with \madanalysis\ with those of the
CMS analysis note. We start from parton-level event samples
that describe the supersymmetric signal in the context of specific benchmark scenarios,
and which have been provided by the CMS collaboration. We then make use
of the {\sc Pythia}~6 program~\cite{Sjostrand:2006za} for parton showering
and hadronization, and of the previously mentioned modified version
of the \textsc{Delphes}~3 package~\cite{deFavereau:2013fsa,Brooijmans:2014eja}
for a simulation of the detector response which
uses the CMS detector description of Ref.~\cite{Agram:2013koa}.
Our final number of events are normalized to a signal cross section
including the merging of
the next-to-leading order result with resummed predictions at the
next-to-leading logarithmic accuracy~\cite{Kramer:2012bx}, $\sigma =
0.0140$~pb, and an
integrated luminosity of 19.5~fb$^{-1}$. More details
on the validation procedure can be found in Refs.~\cite{Dumont:2014tja,Brooijmans:2014eja}.

\renewcommand\arraystretch{1.5}
\begin{table*}
  \centering
  \begin{tabular}{@{\extracolsep{.1cm}}p{.65\textwidth} | c c}
\hline \hline
  Cut & \madanalysis & CMS \\ \hline
  At least one lepton, four jets and 100~GeV
    of missing transverse energy & $31.4$ & $29.7$ \\
  At least one $b$-tagged jet  & $27.1$ & $25.2$ \\
  No extra loosely-isolated lepton or track & $22.5$ & $21.0$ \\
  No hadronic tau & $22.0$ & $20.6$ \\
  Angular separation between the missing momentum and the two
    hardest jets & $18.9$ & $17.8$ \\
  Hadronic top quark reconstruction & $12.7$ & $11.9$ \\
  The transverse mass $M_T$ (defined in the text) is larger than 120~GeV & $10.4$ & $9.6$ \\
  At least 300~GeV of missing transverse energy and
   $M^W_{T2} > 200$~GeV & $5.1$ & $4.2$ \\
\hline \hline
\end{tabular}
\caption{Cut-flow chart for the benchmark point
  $\tilde{t} \to t \tilde{\chi}^0_1 \  (650/50)$ in the signal region
  investigated in Section~\ref{sec:stop}. We present predictions after each
  of the (pre)selection cuts detailed in the text. The small statistical uncertainties
  are omitted for brevity. We present number of events for 19.5~fb$^{-1}$ of
  simulated collisions, after normalizing the total production rate
  to 0.0140~pb.}
\label{tab:cutflowStop}
\end{table*}
\renewcommand\arraystretch{1.0}

As an example, Table~\ref{tab:cutflowStop} shows the cut-flow
chart for a benchmark point denoted by
$\tilde{t} \to t \tilde{\chi}^0_1 \  (650/50)$
(using the naming scheme of the original CMS analysis). In this scenario,
a pair of top squarks whose mass is equal to 650~GeV is produced, and each squark
then decays with a 100\% branching ratio into
a top quark and the lightest neutralino. The mass of the latter is
fixed to 50~GeV. We present both the output of \madanalysis\
and the CMS expectation~\cite{frankkeithprivatecom} and
observe that an agreement at the level of 20\% has been obtained. This order
of magnitude is not surprising as we are comparing
a fast simulation made with {\sc Delphes} to the full simulation
of CMS.

\begin{figure}
\begin{center}
 \includegraphics[width=0.7\columnwidth]{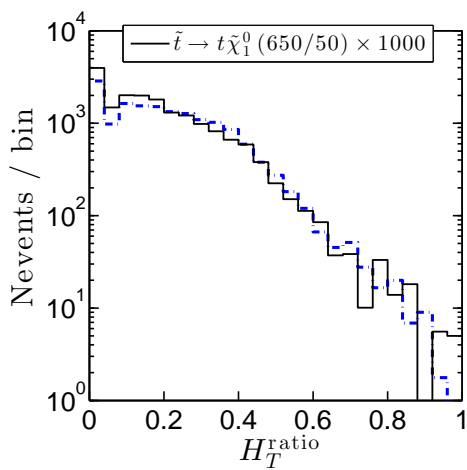}
\end{center}
\caption{Distribution in the $H_T^{\rm ratio}$ variable
  for the \mbox{$\tilde{t} \to t \tilde{\chi}^0_1 \  (650/50)$} scenario
  investigated in Section~\ref{sec:stop}. Only the first four cuts of Table~\ref{tab:cutflowStop}
  have been applied. The solid black line has been obtained with \madanalysis, while the dashed
   blue line is the CMS result.}
\label{fig:htratio}
\end{figure}

The CMS analysis of Ref.~\cite{Chatrchyan:2013xna} contains,
in addition to a cut-based analysis strategy, a second strategy relying on advanced
multivariate techniques. One of the key variables of this analysis
is a quantity denoted by $H_T^{\rm ratio}$,
defined as the fraction of the total scalar
sum of the jet transverse energies (including only jets with transverse momentum
larger than 30~GeV and pseudorapidity $|\eta| < 2.4$) that lies
in the same hemisphere as the missing momentum. For illustrative purposes,
we present below a way to fill a histogram representing this variable.
The definition of $H_T$ appropriate here
being different from Eq.~\eqref{eq:recglobal},
we recalculate it and store
the value in the variable \texttt{HT}.
The amount of hadronic energy located in the same hemisphere
as the missing momentum is stored in the variable \texttt{HTsameHemisphere}.
Both calculations rely on the \texttt{Jets} container, a collection
of relevant jets.
\begin{verbatim}
 double HT = 0.;
 double HTsameHemisphere = 0.;
 for(unsigned int i=0; i<Jets.size(); i++)
 {
   double Et = Jets[i]->momentum().Et();
   HT += Et;
   if(Jets[i]->dphi_0_pi(pTmiss) < 1.5708)
     HTsameHemisphere += Et;
 }
 double HTratio = HTsameHemisphere / HT;
\end{verbatim}

A histogram declared as \texttt{"HTratio"} at the level
of the initialization of the analysis can then
be filled using the syntax introduced in Section~\ref{sec:cutshistos},
\begin{verbatim}
 Manager()->FillHisto("HTratio", HTratio);
\end{verbatim}
The distribution that is extracted from the corresponding output \textsc{Saf}
file is shown in Figure~\ref{fig:htratio},
alongside the CMS result presented in Ref.~\cite{Chatrchyan:2013xna}.
The two are in good agreement.

\subsection{Designing a phenomenological analysis for probing hadronic monotop states}
\label{sec:monotop}

\renewcommand{\arraystretch}{1.5}%
\begin{table*}
  \centering
  \begin{tabular}{@{\extracolsep{.1cm}}p{.55\textwidth} | c c}
  \hline\hline
  Cut & {\bf SII.v-400} & {\bf SII.v-600}\\
  \hline
  Jet selection and lepton veto & 3176 $\pm$ 51.2           & 774 $\pm$ 25.2\\
  Missing energy requirement    & \phantom{0}949 $\pm$ 30.0 & 305 $\pm$ 16.8\\
  $W$-boson reconstructed mass  & \phantom{0}515 $\pm$ 22.4 & 163 $\pm$ 12.5\\
  Separation of the reconstructed top quark from the missing momentum
                                & \phantom{0}501 $\pm$ 22.1 & 158 $\pm$ 12.3\\
  Separation of the hardest jet from the missing momentum
                                & \phantom{0}497 $\pm$ 22.0 & 156 $\pm$ 12.3\\
  Reconstructed top mass        & \phantom{0}311 $\pm$ 17.5 & \phantom{0}96 $\pm$ \phantom{0}9.7\\
  \hline\hline
  \end{tabular}
  \caption{\small \label{tab:monotopcharts}Cut-flow charts for the two considered
   monotop scenarios \mbox{{\bf SII.v-400}} and \mbox{{\bf SII.v-600}} (the
   numbers indicating the choice for the invisible state mass in GeV).
   We present the predicted number of events after each of the cuts
   detailed in the text, for an integrated luminosity of 20~fb$^{-1}$ of LHC collisions
   at a center-of-mass energy of 8~TeV.}
\end{table*}
\renewcommand{\arraystretch}{1}%

The LHC sensitivity to the observation of a monotop state -- a topology
where a single, hadronically decaying top quark is produced in association with
missing energy --
has been recently investigated by means of a phenomenological analysis
relying on a cut-and-count technique~\cite{Agram:2013wda}.
It exploits the presence
of three final-state jets (including a $b$-tagged jet) compatible with a
top quark decay, lying in a different hemisphere to
the (large) missing transverse momentum. In more detail, events are selected
as follows.
\begin{itemize}
 \item Selected events are required to contain two or three light (non-$b$-tagged) jets
   (allowing for one from initial or final state radiation)
   with a transverse momentum
   greater than 30~GeV, as well as one $b$-tagged jet with a transverse
   momentum larger than 50~GeV. The pseudorapidity of each jet must satisfy
   $|\eta^j| < 2.5$ and the ratio of their hadronic to electromagnetic
   calorimetric energy must be above 30\%.
 \item Events featuring isolated charged leptons with a transverse
   momentum $p_T^\ell > 10$~GeV and a pseudorapidity $|\eta^\ell| < 2.5$
   are vetoed. Lepton isolation
   is enforced by imposing that the sum of the transverse momenta of all
   tracks in a cone of $R=0.4$ centered on the lepton is smaller than $0.2 p_T^\ell$.
  \item At least 250~GeV of missing transverse energy is required.
  \item We select the pair of light jets whose invariant mass is the closest
   to the $W$-boson mass. This quantity is then
   constrained to lie in the $[50,105]$~GeV range.
  \item The missing momentum is constrained to be well separated from the momentum
   of the reconstructed top quark (\mbox{$\Delta\phi \in [1,5]$}, the azimuthal angular
   distance being normalized in the $[0,2\pi]$ range).
  \item The missing momentum is constrained to be well separated from the momentum
   of the hardest jet ($\Delta\phi \in [0.5, 5.75]$, the azimuthal angular
   distance being normalized in the $[0,2\pi]$ range).
  \item The reconstructed top mass is constrained to lie in the $[140,195]$~GeV range.
\end{itemize}
As in Section~\ref{sec:stop}, all these cuts can be easily implemented
following the syntax of Section~\ref{sec:ma5}, so that
we again restrict ourselves to the presentation of
a few illustrative samples of code.

First,
we show how to select the jet candidates relevant for the analysis
and store them in a container named \texttt{TheJets},
\begin{verbatim}
 for (unsigned int i=0;
      i<event.rec()->jets.size();i++)
 {
   const RecJetFormat *myj = &(event.rec()->jets[i]);
   double abseta = std::fabs(myj->eta());
   double pt = myj->pt();
   double HEEE  = myj->HEoverEE();
   if(abseta<2.5 && pt > 30 && HEEE>0.3)
     TheJets.push_back(myj);
 }
\end{verbatim}
In those self-explanatory lines of code, we have not yet split the jets
into $b$-tagged and non-$b$-tagged ones.

\begin{figure}
\begin{center}
 \includegraphics[width=0.8\columnwidth]{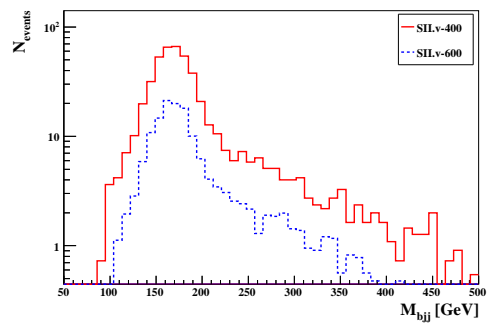}
\end{center}
\caption{Distribution in the reconstructed top mass $M_{bjj}$ for the two new physics scenarios
  investigated in Section~\ref{sec:monotop}, \mbox{{\bf SII.v-400}}
   and \mbox{{\bf SII.v-600}} (the
   numbers indicating the choice for the invisible state mass in GeV). All selection cuts
  except the one on $M_{bjj}$ have been applied.
  The curves are normalized to 20~fb$^{-1}$ of simulated
  LHC collisions at a center-of-mass energy of 8~TeV.}
\label{fig:mtopreco}
\end{figure}

Second, we focus on the reconstruction
of the $W$-boson and the top quark. Assuming that all selected light
jets are stored in a vector named \texttt{ljets} and that the $b$-tagged jet
is represented by the variable \texttt{bjet}, a possible implementation
of a code deriving the four-momenta of the
reconstructed $W$-boson and top quark would be
\begin{verbatim}
 TLorentzVector w, top;
 for(unsigned int i=0; i<ljets.size(); i++)
  for(unsigned int j=i+1; j<ljets.size(); j++)
  {
    TLorentzVector w_tmp =
      ljets[i]->momentum()+ljets[j]->momentum();

    if(i==0 && j==1 ||
       std::fabs(w_tmp.M()-80.)<std::fabs(w.M()-80.))
      w=w_tmp;
  }

  top = w + bjet->momentum();
\end{verbatim}
In the lines above, the double \texttt{for}-loop
derives the pair of light jets that form the system which
is the most compatible with a $W$-boson.
The four-momentum of the reconstructed $W$-boson is saved
as an instance of the \texttt{TLorentzVector} class (named \texttt{w})
and the four-momentum of the reconstructed top quark is then derived by
adding the four-momentum of the
$b$-tagged jet (stored in the \texttt{top} variable).
The reconstructed top mass could then be histogrammed via the
standard command,
\begin{verbatim}
 Manager()->FillHisto("Mtreco", top.M());
\end{verbatim}
where a histogram named \texttt{"Mtreco"}
has been initialized appropriately.
Moreover, the selection cut on this variable could be implemented via
\begin{verbatim}
 bool cutcondition = (top.M()>140) && (top.M()<195);
 if(!Manager()->ApplyCut(cutcondition,"Mtop"))
    return;
\end{verbatim}
assuming that the cut named \texttt{"Mtop"} has been correctly initialized.

We apply the above analysis in the context of the
{\bf SII.v} monotop scenario of Ref.~\cite{Agram:2013wda}.
In this setup, the monotop system arises from the flavor-changing interaction
of an up quark with a novel invisible vector boson whose mass have been
fixed to either 400~GeV or 600~GeV. Using the publicly available
{\sc FeynRules}~\cite{Christensen:2008py,Alloul:2013bka} monotop
model~\cite{Andrea:2011ws}, we generate a UFO
library~\cite{Degrande:2011ua} that we link to the
\textsc{MadGraph}~5 event generator~\cite{Alwall:2011uj} that is used to simulate
parton-level events that include the decay of the top quark. We then perform parton
showering, hadronization and the simulation of the detector response
as in Section~\ref{sec:stop}.
From our analysis implementation, we derive,
in the context of the two considered new physics models,
the distribution in the reconstructed top mass $M_{bjj}$ of
Figure~\ref{fig:mtopreco} and the cut-flow charts of
Table~\ref{tab:monotopcharts}.

\section{Conclusion}
\label{sec:conclusions}
We have presented a major extension of
the expert mode of the \madanalysis\ package. Both designing a prospective
new physics analysis and recasting an experimental search featuring multiple
signal regions can now be achieved in a user-friendly fashion that relies
on a powerful handling of regions, histogramming and selection cuts. We have
illustrated the strengths of our approach with two examples. First,
we have reimplemented a CMS search for stops in events with a single lepton
and missing energy. We have shown that predictions of \madanalysis\ agree
with the CMS expectation at the level of 20\% for a specific
benchmark scenario. Second, we have implemented a phenomenological study
for estimating the LHC sensitivity to hadronically decaying monotop systems
that has been employed in a recent publication.

\section*{Acknowledgments}
We are grateful to S.~Kraml, J.~So and P.~Wymant for helpful discussions
and remarks, as well as to A.~Alloul, J.~Andrea, J.~Bernon, S.~Bein,
M.~Blanke, G.~Chalons, K.~de~Causmaecker, I.~Galon,
M.~Kraft, S.~Kulkarni, A.~Mariotti,
K.~Mawatari, L.~Mitz\-ka, G.~Perez, C.~Petersson, D.~Redigolo,
T.~Schmitt and D.~Sengupta
for their help in testing and debugging the code.
We also thank J.~Alwall, R.~Frederix, F.~Maltoni, O.~Mattelaer and
T.~Stelzer for their support
from the early stages of the \madanalysis\ project. EC, BF and CW
are greatful to the LPSC Grenoble for hospitality.
This work has been partially supported by
the Theory-LHC France initiative of the CNRS/IN2P3, a Ph.D.~fellowship
of the French ministry for education and research and
the French ANR projects
12-BS05-0006 DMAstroLHC and 12-JS05-002-01 BATS@LHC.
\bibliographystyle{spphys}
\bibliography{MyBib}

\end{document}